\documentclass{article}
\usepackage{spconf,amsmath,graphicx}

\usepackage{times}
\usepackage{epsfig}
\usepackage{amssymb}
\usepackage{booktabs}
\usepackage{caption}
\usepackage{multirow}
\usepackage{subcaption}
\usepackage{caption}

\usepackage{enumitem}
\setlist{nosep, leftmargin=14pt}

\usepackage{mwe} 

\title{SIAN: Style-Guided Instance-Adaptive Normalization for Multi-Organ Histopathology Image Synthesis}
\name{Haotian Wang, Min Xian$^{*}$\thanks{$^{*}$ corresponding author. mxian@uidaho.edu}, Aleksandar Vakanski, Bryar Shareef}
\address{Department of Computer Science, University of Idaho, USA}
\begin{document}
%
\maketitle
\begin{abstract}
Existing deep neural networks for histopathology image synthesis cannot generate image styles that align with different organs, and cannot produce accurate boundaries of clustered nuclei. To address these issues, we propose a style-guided instance-adaptive normalization (SIAN) approach to synthesize realistic color distributions and textures for histopathology images from different organs. SIAN contains four phases, semantization, stylization, instantiation, and modulation. The first two phases synthesize image semantics and styles by using semantic maps and learned image style vectors. The instantiation module integrates geometrical and topological information and generates accurate nuclei boundaries. We validate the proposed approach on a multiple-organ dataset, Extensive experimental results demonstrate that the proposed method generates more realistic histopathology images than four state-of-the-art approaches for five organs. By incorporating synthetic images from the proposed approach to model training, an instance segmentation network can achieve state-of-the-art performance.  

\end{abstract}
\section{Introduction}
\label{sec:intro}

Histopathology image analysis has achieved great success in automatic tissue segmentation \cite{hover, wang2020bending} and cancer grading \cite{shaban2020context}. Existing deep learning-based methods require large fully-annotated datasets during the training stage, but current annotated datasets are relatively small. For example, only tens of image patches were used in \cite{hover, MoNuseg, vu2019methods, naylor2018segmentation}. With large annotated datasets, we could train more accurate and reliable models. However, it is expensive to annotate large datasets for histopathology images, because each image may contain more than tens of thousands of nuclei. 

\begin{figure}
\begin{center}
    \small
      \begin{subfigure}[b]{0.155\linewidth}
        \captionsetup{type=figure, labelformat=empty}
        \caption{Breast}
        \vspace*{-1mm}  
        \includegraphics[width=\linewidth]{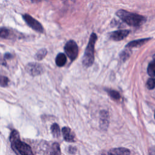}
      \end{subfigure}
      \begin{subfigure}[b]{0.155\linewidth}  
        \captionsetup{labelformat=empty}
        \caption{Prostate}
        \vspace*{-1mm}  
        \includegraphics[width=\linewidth]{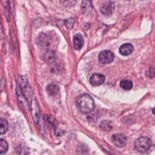}
      \end{subfigure}
      \begin{subfigure}[b]{0.155\linewidth} 
        \captionsetup{labelformat=empty}
        \caption{Liver}
        \vspace*{-1mm}  
        \includegraphics[width=\linewidth]{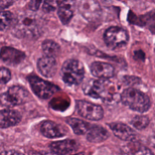}
      \end{subfigure}  
      \begin{subfigure}[b]{0.155\linewidth}
        \captionsetup{labelformat=empty}
        \caption{Bladder}
        \vspace*{-1mm}
        \includegraphics[width=\linewidth]{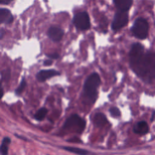}
      \end{subfigure}
      \begin{subfigure}[b]{0.155\linewidth}   
        \captionsetup{labelformat=empty}
        \caption{Kidley}
        \vspace*{-1mm}    
        \includegraphics[width=\linewidth]{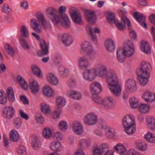}
      \end{subfigure}
      \begin{subfigure}[b]{0.155\linewidth}   
        \captionsetup{labelformat=empty}
        \caption{Colon}
        \vspace*{-1mm}    
        \includegraphics[width=\linewidth]{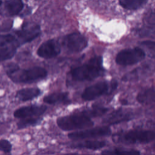}
      \end{subfigure} 

      \begin{subfigure}[b]{0.155\linewidth}
        \includegraphics[width=\linewidth]{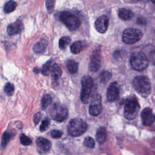}
      \end{subfigure}
      \begin{subfigure}[b]{0.155\linewidth}  
        \includegraphics[width=\linewidth]{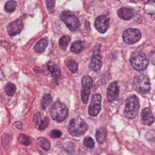}
      \end{subfigure}
      \begin{subfigure}[b]{0.155\linewidth} 
        \includegraphics[width=\linewidth]{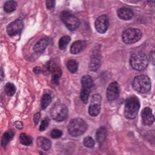}
      \end{subfigure}  
      \begin{subfigure}[b]{0.155\linewidth}
        \includegraphics[width=\linewidth]{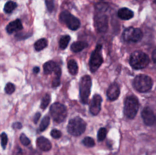}
      \end{subfigure}
      \begin{subfigure}[b]{0.155\linewidth}   
        \includegraphics[width=\linewidth]{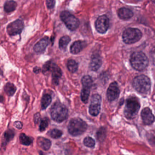}
      \end{subfigure}
      \begin{subfigure}[b]{0.155\linewidth}   
        \includegraphics[width=\linewidth]{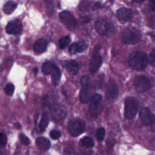}
      \end{subfigure}

      \begin{subfigure}[b]{0.155\linewidth}
        \includegraphics[width=\linewidth]{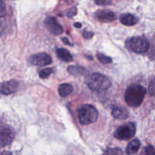}
      \end{subfigure}
      \begin{subfigure}[b]{0.155\linewidth}  
        \includegraphics[width=\linewidth]{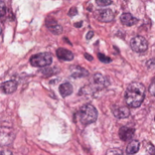}
      \end{subfigure}
      \begin{subfigure}[b]{0.155\linewidth} 
        \includegraphics[width=\linewidth]{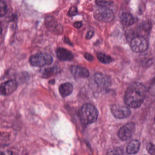}
      \end{subfigure}  
      \begin{subfigure}[b]{0.155\linewidth}
        \includegraphics[width=\linewidth]{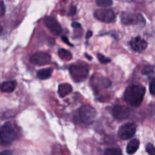}
      \end{subfigure}
      \begin{subfigure}[b]{0.155\linewidth}   
        \includegraphics[width=\linewidth]{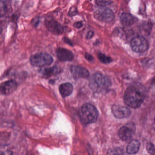}
      \end{subfigure}
      \begin{subfigure}[b]{0.155\linewidth}   
        \includegraphics[width=\linewidth]{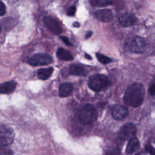}
      \end{subfigure}

 \end{center}
    \vspace*{-6mm}
   \caption{Examples of image synthesis for multiple organs. The top row shows real histopathology images from six organs. The second and third rows are synthesized images generated using the proposed approach (SIAN).}
\label{fig:long}
\label{fig:onecol}
\end{figure}

To overcome the challenge, image synthesis is adopted. Recent works have demonstrated that high-quality synthetic images could improve the overall performance in histopathology image analysis \cite{deshpande2022safron, wei2019generative, xue2021selective}. However, these methods generated images only for single cancer or cancers with shared similarity, e.g., colorectal cancer \cite{deshpande2022safron, wei2019generative}, lymph node \cite{xue2021selective}; thus their models cannot generate different image styles for different cancer types. In practice, H\&E-stained images for cancers from different organs could have large color and texture variances both in foreground nuclei and background stroma (first row of Fig. 1). Therefore, it is important to have the network to generate histopathology images well in various stain distribution across multiple organs. 
    
Recently, neural style transfer (NSF) methods have been widely exploited in many natural image synthesis tasks for manipulating image styles \cite{huang2017arbitrary, SPADE, karras2019style}, they aim to learn the style from a reference image and apply it to the target image. SPADE \cite{SPADE} extended the AdaIN norm \cite{huang2017arbitrary} into the spatially-adaptive manner for obtaining semantic alignments and used the encoded style vector at the beginning of a network, which enabled simultaneously style manipulation and semantic image synthesis. However, most existing histopathology image synthesis methods applied semantic layouts as the network input to learn object-level image appearance \cite{SPADE, deshpande2022safron, wei2019generative}. In histopathology image analysis, a large amount of clustered and overlapped objects may have the same semantic class label, which makes it difficult to generate accurate boundaries among clustered objects.

\begin{figure*}[t]
\begin{center}
   \includegraphics[width=0.8\linewidth]{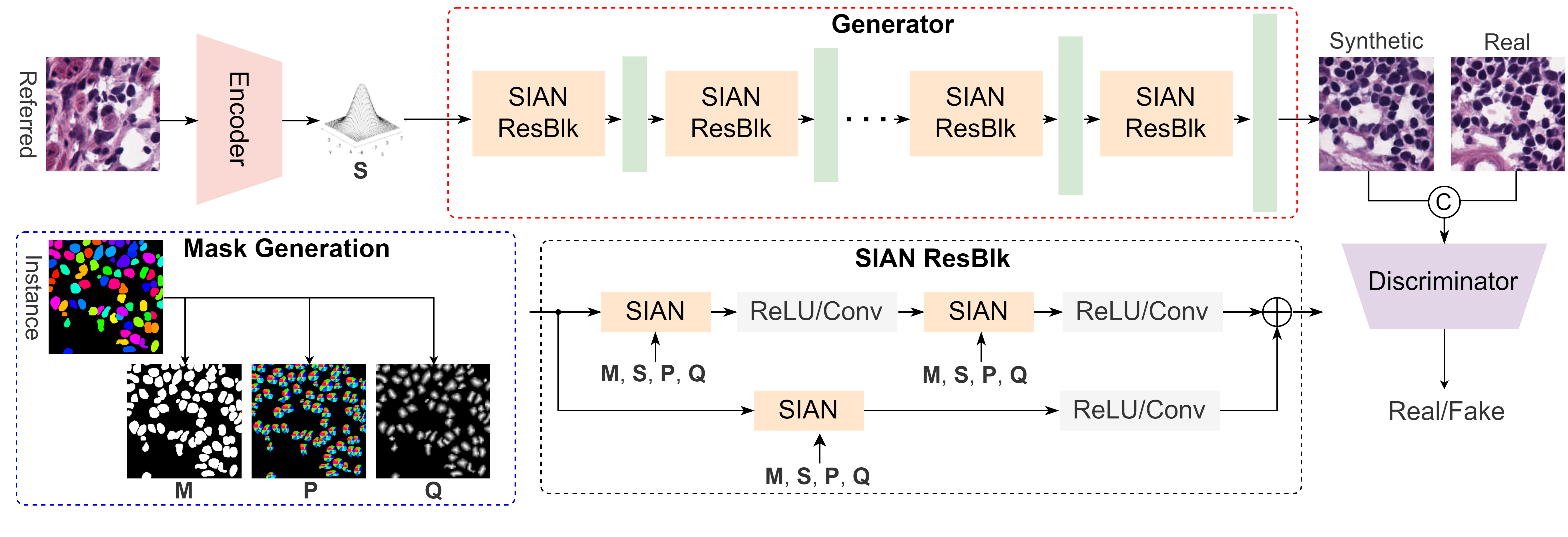}
\end{center}
    \vspace*{-9.2mm}
   \caption{Architecture of the proposed method. The encoder learns style vectors from a referred image; and the SIAN blocks integrate image style ($\mathbf{S}$), semantic map ($\mathbf{M}$), directional map ($\mathbf{P}$), and distance map ($\mathbf{Q}$) into a generator network.}
\end{figure*}

To alleviate the above issues, we proposed a style-guided instance-adaptive normalization (SIAN) to combine image style vector with instance layout for modulating the GAN generator. The learned transformation can effectively propagate the network to learn style factors for synthesizing histopathology images across various color distributions; the captured instance features can preserve geometrical and topological information for generating accurate densely-clustered nuclei. SIAN can generate style images from a specific organ and synthesize histopathology images with a similar style.  

\section{Proposed Method}
\label{sec:method}

\subsection{Architecture and learning objectives}
Fig. 2 shows the overall architecture of the proposed generator. The proposed generator has seven SIAN residual blocks (SIAN ResBlk), and each SIAN ResBlk is followed by a up-sampling layer. Each SIAN ResBlk contains two consecutive SIAN blocks, and each is followed by ReLU and convolutional layers. The skip connection has a SIAN block, a ReLU, and a convolutional layer. All input maps are down-sampled to the same height and width with the corresponding feature maps in the generator. We follow the same encoder and discriminator architectures described in \cite{SPADE}.  The overall loss function contains five loss components, and is defined by \cite{SPADE}
\begin{equation}
\mathcal{L}_{SIAN}=\mathcal{L}_{GAN}+\lambda_{1}\mathcal{L}_{F}+\lambda_{2}\mathcal{L}_{P} + \\
\lambda_{3}\mathcal{L}_{KLD} + \lambda_{4}\mathcal{L}_{reg}
\end{equation}
where $\mathcal{L}_{GAN}$ is the hinge-based conditional adversarial loss \cite{spectral}, $\mathcal{L}_{F}$ is the feature matching loss in the multi-scale discriminator \cite{wang2018high}, $\mathcal{L}_{P}$ is the perceptual loss \cite{johnson2016perceptual} for minimizing the features between real and synthetic images, and $\mathcal{L}_{KLD}$ is the KL divergence loss \cite{KingmaW13} for the encoder to constrain the style vector to the standard Gaussian distribution. $\lambda_{1}$, $\lambda_{2}$, $\lambda_{3}$, and $\lambda_{4}$ controls the contributions of different loss terms.

\subsection{Style-guided instance-adaptive normalization}
We propose a new conditional normalization block, namely, the Style-guided Instance-Adaptive Normalization (SIAN) to learn instance-level features and integrate image styles for cancers from different organs. Fig. 3 shows the details of the SIAN block. The block has four phases: semantization, stylization, instantiation, and modulation. The block takes four inputs besides image feature maps, i.e., semantic mask $\mathbf{M}$, style vector $\mathbf{S}$, direction mask $\mathbf{P}$, and distance mask $\mathbf{Q}$. The semantization phase embeds image semantics from the input mask; the stylization creates a style matrix from a referred image and integrates image semantics and style. The instantiation phase uses direction and distance maps to distinguish individual nuclei. The modulation phase learns the scale and bias and integrates them into the network.     

\begin{figure}
\begin{center}
   \includegraphics[width=0.95\linewidth]{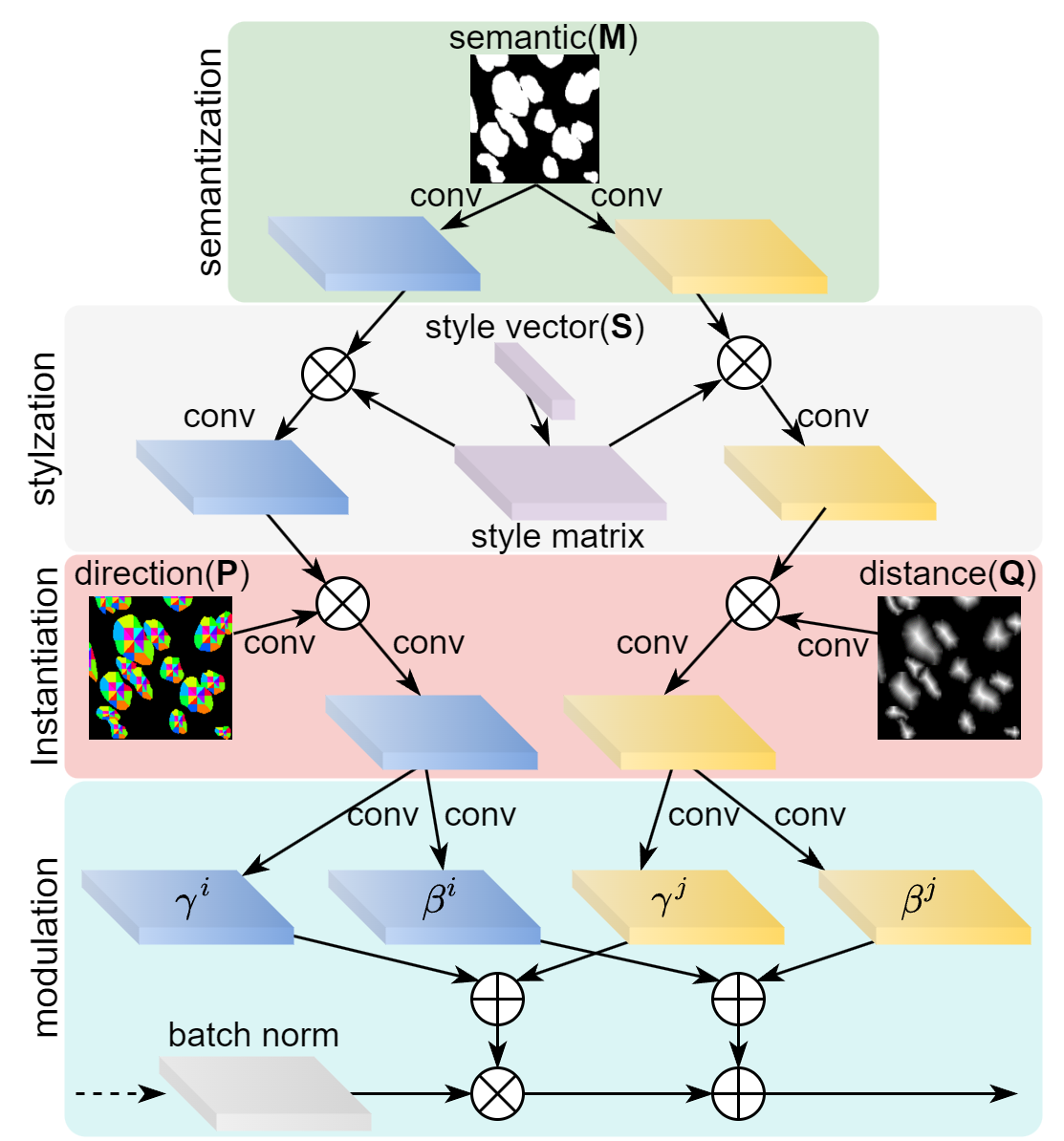}
\end{center}
   \vspace*{-7.2mm}
   \caption{SIAN normalization. SIAN block takes four inputs: semantic masks $\mathbf{M}$, style vector $\mathbf{S}$, directional mask $\mathbf{P}$, and distance mask $\mathbf{Q}$ for combining different features at multiple phases in the block. $\otimes$ denotes element-wise multiplication, and $\oplus$ is element-wise addition.}
\end{figure}

Let $\mathbf{h}$ denote the input activation of the current layer of the proposed neural network with a batch size of $N$. Let $H$, $W$ and $C$ denote the height, width, and channels of an activation map in $i$th layer. The final modulated activation value ($n \in N$, $c \in C$, $y \in H$, $x \in W$) is defined as 
\begin{equation}
\gamma_{c,y,x}(\mathbf{M},\mathbf{S},\mathbf{P},\mathbf{Q})\frac{h_{n,c,y,x}-\mu_{c}}{\sigma_{c}} + \beta_{c,y,x}(\mathbf{M},\mathbf{S},\mathbf{P},\mathbf{Q})
\vspace*{-1mm}
\end{equation}
where $h_{n,c,y,x}$ is the activation output before normalization; the modulation parameter $\gamma_{c,y,x}$ and $\beta_{c,y,x}$ are the element-wise summation of modulation parameters of two branches, i.e.,  $\gamma_{c,y,x}^{i} + \gamma_{c,y,x}^{j}$ and $\beta_{c,y,x}^{i} + \beta_{c,y,x}^{j}$. $\mu_{c}$ and $\sigma_{c}$ are the mean and standard deviation of the activation of the channel $c$, respectively. In the SIAN block, the semantic layout first passes to two convolutional layers, which split the semantic information into two separate branches to learn the directional features and distance features separately. The two branches have the same architecture. In each branch, the convolutional kernel first  multiplies with the reshaped style vector, which combines  style factors in the block. After that, the instance layouts (direction or distance) are fed through a 1 $\times$ 1 convolutional layer and multiplied with the previous convolutional layer. The next convolutional layer learns the compensation of semantic, style, and instance features and then split into two convolutional layers to learn the modulation parameters ($\gamma$ and $\beta$) spatially. Finally, those modulation parameters and the output of batch normalization are integrated for accurate histopathology image synthesis. All convolutional layers in SIAN use the 3$\times$3 kernel size with 128 filters. 

Instance masks are applied to generate the semantic mask, and nuclei directional and distance maps. The semantic map is used to separate nuclei and stroma, and the directional and distance maps are useful to demonstrate the boundaries and centroids between two or more touching nuclei. We employed the 2-bin direction mask \cite{masklab} and Medial Axis (MA) distance mask \cite{tanet} as the instance descriptors. Direction map provides important centroid and directional information of nuclei. MA distance mask shows the distance between the nucleus boundary to its skeleton while providing nuclei topological and geometrical features. 


\section{Experimental Results}

\subsection{Dataset, metrics, and setting}

\textbf{Dataset.} The experiments are conducted on the multi-organ nuclei segmentation dataset (MoNuSeg) \cite{MoNuseg} which has 44 H\&E stained histopathology image patches. Both the training and test sets contain images from six organs including breast, liver, kidney, prostate, bladder, and colon; while the training set includes stomach as the seventh organ, and the testing set has brain images. 

\textbf{Evaluation metrics.} We employed five metrics to evaluate the method performance for image synthesis, e.g., FID \cite{heusel2017gans}, SSIM \cite{wang2004image}, DQ \cite{panoptic}, SQ \cite{panoptic}, and PQ \cite{panoptic}. We used two metrics FID and SSIM to measure the distribution distance and structural similarity between real images and synthetic images, respectively; and used DQ, SQ, and PQ are utilized to assess the nuclei segmentation performance. Specifically, we run a pre-trained segmentation model (SegNet \cite{segnet}) which is trained on real images, and then test and evaluate using the synthetic images. In addition, we show the visual comparison of our synthetic images compared to other methods.

\textbf{Implementation details.} The input image size of our approach is 256$\times$256. We used random flip, rotation and median blur for data augmentation. We use the ADAM optimizer with the total training epochs of 50 and batch size of 8 to train the network. The experiments are conducted on a NVIDIA RTX 8000 GPU. 

During inference, for the style encoder, we take an arbitrary histopathology image as input and output the encoded style vectors. Then, the encoded style vectors together with an arbitrary instance mask are input into the trained generator network  to produce histopathology images.

\subsection{Image quality assessment}
The proposed method is compared with four state-of-the-art image synthesis models: pix2pix GAN \cite{pix2pix}, Sharp-GAN \cite{sharpgan}, pix2pixHD \cite{pix2pixhd}, and SPADE \cite{SPADE} using FID, SSIM, DQ, SQ, and PQ metrics. The quantitative results of different approaches on the MoNuSeg test set are shown in Table 1. The proposed method outperforms the state-of-the-art methods both in image reconstruction quality using SSIM and FID, and segmentation quality using PQ, SQ and DQ. In addition, we integrated instantiation phase (INST) only, and the SIAN block with the style vectors (STYLE), all the evaluation metrics are improved from the baseline SPADE. Overall, we can conclude that our SIAN achieved the best quantitative performance among other methods. 

\begin{figure}
\begin{center}
  \begin{subfigure}[b]{0.18\linewidth}
    \vspace*{-2.5mm}
    \includegraphics[width=\linewidth]{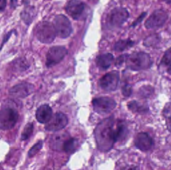}
  \end{subfigure}
  \begin{subfigure}[b]{0.18\linewidth}
    \vspace*{-2.5mm}    
    \includegraphics[width=\linewidth]{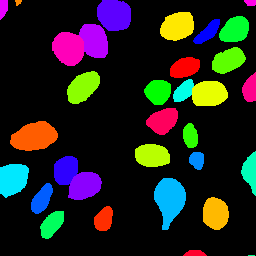}
  \end{subfigure}
  \begin{subfigure}[b]{0.18\linewidth}
    \vspace*{-2.5mm}    
    \includegraphics[width=\linewidth]{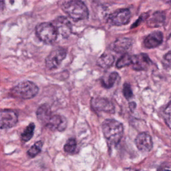}
  \end{subfigure}
  \begin{subfigure}[b]{0.18\linewidth}
    \vspace*{-2.5mm}    
    \includegraphics[width=\linewidth]{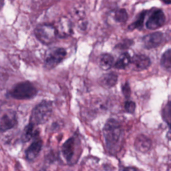}
  \end{subfigure}
  \begin{subfigure}[b]{0.18\linewidth}
    \vspace*{-2.5mm}    
    \includegraphics[width=\linewidth]{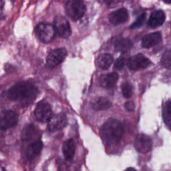}
  \end{subfigure}

  \begin{subfigure}[b]{0.18\linewidth}
    \includegraphics[width=\linewidth]{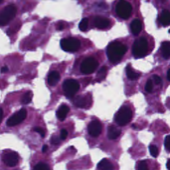}
  \end{subfigure}
  \begin{subfigure}[b]{0.18\linewidth}
    \includegraphics[width=\linewidth]{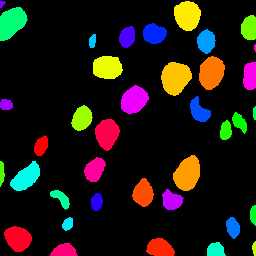}
  \end{subfigure}
  \begin{subfigure}[b]{0.18\linewidth}
    \includegraphics[width=\linewidth]{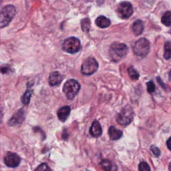}
  \end{subfigure}
  \begin{subfigure}[b]{0.18\linewidth}
    \includegraphics[width=\linewidth]{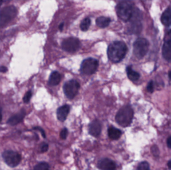}
  \end{subfigure}
  \begin{subfigure}[b]{0.18\linewidth}
    \includegraphics[width=\linewidth]{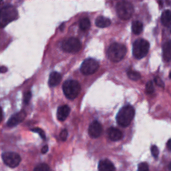}
  \end{subfigure}

  \begin{subfigure}[b]{0.18\linewidth}
    \includegraphics[width=\linewidth]{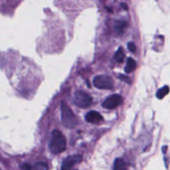}
  \end{subfigure}
  \begin{subfigure}[b]{0.18\linewidth}
    \includegraphics[width=\linewidth]{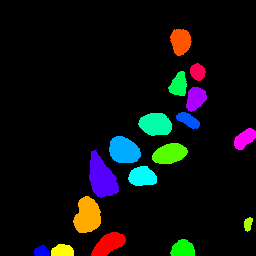}
  \end{subfigure}
  \begin{subfigure}[b]{0.18\linewidth}
    \includegraphics[width=\linewidth]{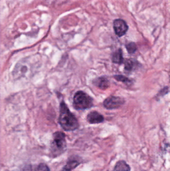}
  \end{subfigure}
  \begin{subfigure}[b]{0.18\linewidth}
    \includegraphics[width=\linewidth]{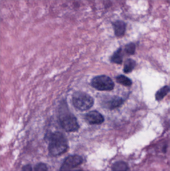}
  \end{subfigure}
  \begin{subfigure}[b]{0.18\linewidth}
    \includegraphics[width=\linewidth]{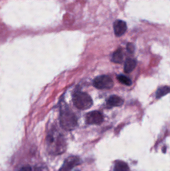}
  \end{subfigure}
  
\end{center}
    \vspace*{-6mm}
   \caption{Visual comparison of histopathology image synthesis for the MoNuSeg test set. (a) Image patches, (b) nuclei masks, (c-e) results of Sharp-GAN, SPADE, and ours, respectively}
\label{fig:long}
\label{fig:onecol}
\end{figure}

\begin{table}
\small
\begin{center}
\begin{tabular}{l||c|c|c|c|c}
\hline
Methods & FID$\downarrow$ & SSIM$\uparrow$ & DQ$\uparrow$ & SQ$\uparrow$ & PQ$\uparrow$ \\
\hline
pix2pix \cite{pix2pix} & 170.1 & 0.467 & 0.687 & 0.717 & 0.493\\
Sharp-GAN \cite{sharpgan} & 155.2 & 0.483 & 0.721 & 0.745 & 0.538\\
pix2pixHD \cite{pix2pixhd} & 186.3 & 0.479 & 0.750 & 0.753 & 0.566 \\
SPADE \cite{SPADE} & 134.6 & 0.488 & 0.705 & 0.738 & 0.552\\

INST & 125.4 & 0.491 & 0.748 & 0.768 & 0.575 \\
STYLE & 116.5 & 0.506  & 0.743 & 0.769 &  0.571\\

SIAN & \textbf{115.7} & \textbf{0.515} & \textbf{0.757} & \textbf{0.761} & \textbf{0.586}\\
\hline  
\end{tabular}
\end{center}
\vspace*{-6mm}
\caption{Overall performance on MoNuSeg datasets with reconstruction metrics and segmentation metrics.}
\end{table}

Fig. 4 compares the proposed method, Sharp-GAN \cite{sharpgan}, and SPADE \cite{SPADE} using three examples. We noted that Sharp-GAN cannot recover the texture and color distributions of nuclei and stroma in real images, especially in the first and second row. I.e., the synthetic nuclei have different appearances from real nuclei, and their background stroma lacks meaningful texture and color. SPADE achieved better performance compared to Sharp-GAN, but the generated images were not realistic. Our SIAN generates more realistic images than SPADE and Sharp-GAN. SPADE used the semantic layout as input, while our method used the instance layouts. As shown in Figure 5, SPADE tends to generate blur and incorrect nuclei in the clustered region. Our approach produces more accurate boundaries for clustered nuclei. 

\begin{figure}
\begin{center}

  \begin{subfigure}[b]{0.16\linewidth}
    \includegraphics[width=\linewidth]{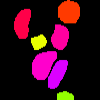}
  \end{subfigure}
  \begin{subfigure}[b]{0.16\linewidth}
    \includegraphics[width=\linewidth]{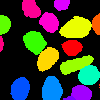}
  \end{subfigure}  
  \begin{subfigure}[b]{0.16\linewidth}
    \includegraphics[width=\linewidth]{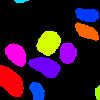}
  \end{subfigure}
  \begin{subfigure}[b]{0.16\linewidth}
    \includegraphics[width=\linewidth]{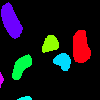}
  \end{subfigure}
  \begin{subfigure}[b]{0.16\linewidth}
    \includegraphics[width=\linewidth]{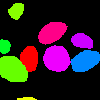}
  \end{subfigure}

  \begin{subfigure}[b]{0.16\linewidth}
    \includegraphics[width=\linewidth]{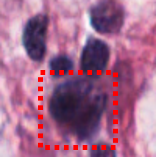}
  \end{subfigure}
  \begin{subfigure}[b]{0.16\linewidth}
    \includegraphics[width=\linewidth]{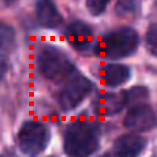}
  \end{subfigure}  
  \begin{subfigure}[b]{0.16\linewidth}
    \includegraphics[width=\linewidth]{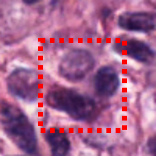}
  \end{subfigure}
  \begin{subfigure}[b]{0.16\linewidth}
    \includegraphics[width=\linewidth]{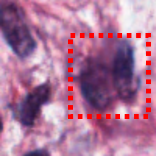}
  \end{subfigure}
  \begin{subfigure}[b]{0.16\linewidth}
    \includegraphics[width=\linewidth]{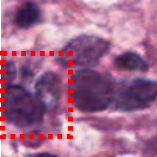}
  \end{subfigure}

  \begin{subfigure}[b]{0.16\linewidth}
    \includegraphics[width=\linewidth]{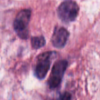}
  \end{subfigure}
  \begin{subfigure}[b]{0.16\linewidth}
    \includegraphics[width=\linewidth]{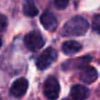}
  \end{subfigure}  
  \begin{subfigure}[b]{0.16\linewidth}
    \includegraphics[width=\linewidth]{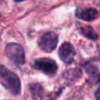}
  \end{subfigure}
  \begin{subfigure}[b]{0.16\linewidth}
    \includegraphics[width=\linewidth]{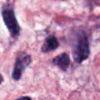}
  \end{subfigure}
  \begin{subfigure}[b]{0.16\linewidth}
    \includegraphics[width=\linewidth]{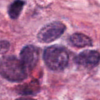}
  \end{subfigure}

\end{center}
    \vspace*{-6mm}
   \caption{Synthesis for clustered nuclei. First row: instance masks (different colors represent different nuclei). Second row: results of SPADE \cite{SPADE}. Third row: results of SIAN.}
\label{fig:long}
\label{fig:onecol}
\end{figure}

\begin{table}
\footnotesize
\begin{center}
\begin{tabular}{l||c|c|c|c|c|c|c}
\hline
Meth. & Bre. & Kid. & Pro. & %
    Bla. & Col. & Lun. & Bra. \\
\hline
\cite{pix2pix} & 211.0 & 197.3 & 237.5 & 246.9 & 239.3 & \textbf{187.5} & 304.4 \\
\cite{sharpgan} & \textbf{198.0} & 190.4 & 235.3 & 211.4& 208.6 & 195.4 & 260.3\\
\cite{pix2pixhd} & 260.0 & 222.1 & 260.1 & 278.6 & 221.8 & 252.8 & 278.6\\
\cite{SPADE} & 212.4 & 187.9 & 200.4 & 212.5 & 222.3 & 207.2 & 245.5\\
SIAN & 200.4 & \textbf{173.9} & \textbf{190.2} & \textbf{196.9} & \textbf{204.1} & 194.6 & \textbf{239.6}\\
\hline  
\end{tabular}
\end{center}
\vspace*{-6mm}
\caption{Performance comparison of image synthesis for multiple organs using the FID score.}
\end{table}

\subsection{Multi-organ image synthesis}
To evaluate synthetic images across multiple organs, we compare the generation performance of four methods for each organ using FID scores. The results are shown in Table 2. The proposed method outperforms other state-of-the-art methods in kidney, prostate, bladder, colon, and brain images, and achieved the second-best results for synthesizing breast and lung images. Fig. 1 shows the results of SIAN across multiple organs, the color and texture distribution of foreground nuclei and background stroma are close to the reals.
  
\subsection{Nuclei segmentation using synthetic images}
In this experiment, we evaluate the effectiveness of synthetic augmentation for training segmentation networks. We train SegNet \cite{segnet} with different input configurations (as shown in Table 3). In experiments, nucleus-like polygons are generated as the synthetic nuclei instance masks \cite{hou2019robust}, in total 5,000 synthetic instance masks are generated and applied to produce corresponding semantic, directional, and distance masks. Then, the pre-trained SIAN is used to apply seven different style vectors encoded from seven different organs (around 700 synthetic images per organ) and generate realistic histopathology images. Finally, we test and evaluate the segmentation performance with the MoNuSeg test set using DQ, SQ, PQ metrics. We compared the proposed approach to other methods on synthetic augmentation. Synthetic images generated from other methods follow their design. As shown in Table 3, with synthetic training images from pix2pixHD, Sharp-GAN, and SPADE, the performance of nuclei segmentation could be significantly improved. The proposed SIAN help generate the best segmentation performance.

\begin{table}[t]
\small
\begin{center}
\begin{tabular}{l||c|c|c}
\hline
Training Set & DQ & SQ & PQ \\
\hline
MoNuSeg training set & 0.704 & 0.737 & 0.521 \\  
MoNuSeg training set$^{*}$ & 0.732 & 0.739 & 0.538\\
+pix2pixHD & 0.742 & 0.737  & 0.544 \\ 
+Sharp-GAN & 0.740 & 0.739 & 0.547\\
+SPADE & 0.743 & 0.738 & 0.549 \\
+SIAN & \textbf{0.748} & \textbf{0.742} & \textbf{0.555}\\
\hline  
\end{tabular}
\end{center}
\vspace*{-6mm}
\caption{Performance of SegNet using different training sets. $^{*}$ denotes the training set augmented using traditional augmentation techniques, e.g., flip, rotate, blur. '+method' denotes applying synthetic augmentation$^{*}$ with 5,000 synthetic images generated from 'method' to the training set.}
\end{table}

\section{Conclusion}

In this paper, we propose the style-guided instance-adaptive normalization (SIAN) for multi-organ histopathology image synthesis, which integrates instance layouts and encodes style vectors into a generative network. SIAN synthesizes histopathology images with styles that align with the image styles of multiple organs. SIAN utilizes the directional and distance masks from the nuclei instance maps and generates clear boundaries for densely-clustered nuclei. With the integration of the stylization phase, SIAN allows style editing for synthesizing images of multiple organs. In addition, SIAN demonstrates its effectiveness in augmenting the training set and improving the overall performance of a deep learning model for nuclei segmentation. 

{\small
\bibliographystyle{IEEEbib}
\bibliography{strings,refs}
}
\end{document}